\documentclass[aps,pre,twocolumn,groupedaddress]{revtex4}
\usepackage{graphicx}
\usepackage{bm}
\usepackage{amsmath, amsthm, amssymb}
\usepackage{eurosym}
\usepackage{tabularx}

\begin{document}

\title{Fractional calculus ties the microscopic and macroscopic scales of complex network dynamics}
\author{B.J. West$^{1,2}$}
\author{M. Turalska$^{1}$}
\author{P. Grigolini$^{3}$}
\affiliation{
$^{1}$ Physics Department, Duke University, Durham, NC 27709, USA\\
$^{2}$ Information Science Directorate, US Army Research Office, Research
Triangle Park, NC 27708, USA\\
$^{3}$ Center for Nonlinear Science, University of North Texas, Denton, TX 76203, USA
}
\date{\today }

\begin{abstract}
A two-state master equation based decision making model has been shown to generate
phase transitions, to be topologically complex and to manifest temporal
complexity through an inverse power-law probability distribution function in the switching times between the two critical states of consensus. These properties are entailed by the fundamental assumption that the network elements in the decision making model imperfectly imitate one another. The process of subordination establishes that a single network element can be described by a fractional master equation whose analytic solution yields the observed inverse power-law probability distribution obtained by numerical integration of the two-state master equation to a high degree of accuracy.
\end{abstract}

\maketitle

\section{Introduction}

The extraordinary advancements made in the physical sciences at the beginning of twentieth century originated from the effort to understand the simplest, that is, the most fundamental, elements of physical phenomena. The separation of matter and light into its basic constituents: electrons, protons, atoms, and molecules in the first case and photons in the second, enabled scientists to explain the puzzling properties of solid materials, such as their sound transmission and electrical properties, as well as, light emission and reflection characteristics. The discipline of statistical mechanics demonstrated that the behavior of systems composed of millions of individual particles can be captured with simple laws, involving only their average properties \cite{pauli73}. At that time it seemed as if the Aristotelian approach of reductionism, where complex phenomena consisted of nothing more than the sum of essential elements, was the prescription underlying the correct method of scientific discovery.

The French mathematician Poincar\'{e} was probably the first to rigorously demonstrate the failure of this attractive but overly optimistic method by extending Newton's law of universal gravitation to a system consisting of three celestial bodies \cite{poincare92}. His mathematical treatment of the three-body problem demonstrated that unlike the two-body problem of the earth and sun, a planet's orbit in the three-body system need not be periodic. Poincar\'{e} proved that the long held belief that planetary motion could be built up from the superposition of simple cycles was false. This unanticipated behavior emerges from nonlinear dynamics and has come to be known as chaos theory.

Another perplexing phenomena that violates the principle of superposition is critical phase transitions in magnetic materials. If criticality were truly nothing more than collective behavior resulting from the superposition of some basic building blocks, why then does the change in an external parameter, for example temperature, induce such a dramatic shift in its macroscopic behavior?

More recently the identification of emergent phenomena across multiple disciplines, from the swarming of insects \cite{yates09}, the schooling of fish \cite{katz11} and the flocking of birds \cite{cavagna10} observed in animal groups by naturalists; to the spatiotemporal activity of the brain \cite{beggs03,chialvo10,fraiman09} observed by neurophysiologists; to the collective and cooperative behavior observed in social groups studied by psychologists and sociologists; all demonstrate collective behavior reminiscent of particle dynamics near the critical phase transitions studied by physicists \cite{stanley71}. Each of these disciplines has demonstrated the need to investigate the dynamics of complex networks across scales in order to develop a deeper understanding of how large-scale behavior emerges from microscale dynamics and the sensitivity of the observed behavior to those dynamics.

\begin{figure}[t]
\centering
\includegraphics[scale=0.40]{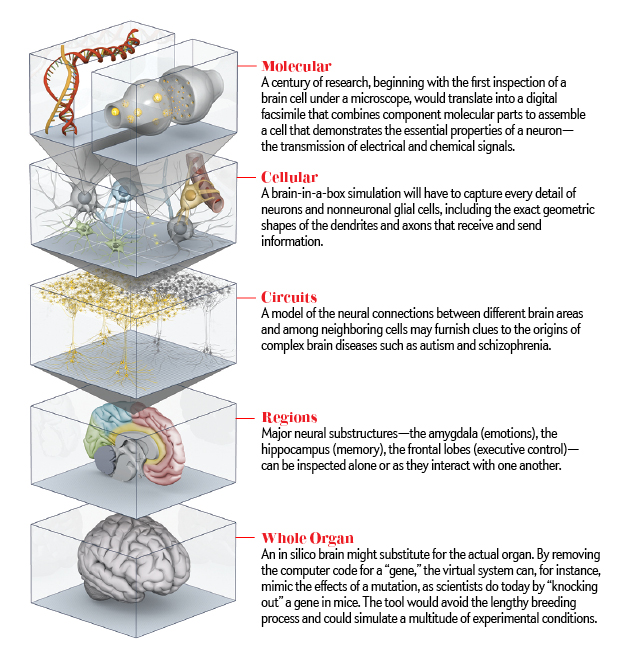}
\caption{ A schematic of the various orders of organisation of the human brain is depicted, starting at the molecular level and ending with the whole organ. Copyright by E. Cooper.}
\label{fig_markram}
\end{figure}

Of particular interest to us here are the biological fields in which we observe a need for a system wide approach \cite{richardson10}. The recent discoveries in biology were propelled by the successes of molecular biology and genetics that have made available genomic blueprints of numerous organisms, which are complemented by extensive experimental data describing cell functions. At the same time however the realization came that biological function emerges out of the interaction of numerous molecular components, such as depicted in Figure \ref{fig_markram} for the human brain, making the detailed knowledge of specific components at any level of organization insufficient to capture macroscopic functionality. There is probably no better example of this limitation then the study of the neurological systems, whose goal is to understand, predict and ultimately modify (in order to heal) brain function. Ongoing initiatives of the Human
Connectome Project \cite{human09}, the Human Brain Project \cite{human13} or the Allen Brain Atlas \cite{human03} illustrate the fact that the system wide approach and integration of data from across different spatial and temporal scales has become the norm in modern scientific disciplines.

Despite experimental developments, the ability of science to make theoretical predictions of the behavior of complex networks is still in its infancy. The adoption of methods from non-equilibrium statistical physics have demonstrated limitations, resulting from the fact that living systems, in contrast to inert physical materials, are extremely heterogeneous, non-generic, highly specialized and operate far from an equilibrium state \cite{elsasser81}. Herein we demonstrate that what was for a very long time a niche branch of mathematics, the fractional calculus, might very well be able to span the gap between the inert materials of physics and the living networks of biology.

Although developed along side the classical calculus, fractional differential equations have only recently been shown to be a convenient way to describe the dynamics of complex phenomena characterized by long-term memory and spatial heterogeneity \cite{magin06, podlubny99, west03}. Fractional differential equations were demonstrated to capture the time
evolution of fractal processes, such as in anomalous diffusion, viscoelasticity and turbulent fluid flow, as reviewed by West and Grigolini \cite{west11}. In spite of the success of the mathematical descriptions of such processes there has been a lack of identification and interpretation of mechanisms that entail fractional dynamic equations in the context of complex networks. Herein we provide an explanation for one source of a fractional differential equation that describes the dynamics of a complex network using a fractional master equation.

Here the utility of the fractional calculus is demonstrated by capturing the dynamics of the individual elements of a complex network from the information quantifying that network's global behavior. The phase transitions observed in complex social and physiologic networks suggest the wisdom of using a generic model from the Ising universality class to characterize network dynamics. Using such a model it is then possible to demonstrate that the individual trajectory response to the collective motion of the network is described by a linear fractional differential equation. The solution to this fractional equation is obtained through a subordination procedure without the necessity of linearizing the underlying dynamics, that is, the solution retains the influence of the nonlinear network dynamics on the individual. Moreover the solutions to the fractional equations of motion suggest a new direction for designing control mechanisms for complex networks.

In Section \ref{DMM} we consider the dynamics of a complex network described by a two-state master equation. The decision making model (DMM), defined by the two-state master equation, undergoes phase transitions at a critical value of the control parameter \cite{turalska11}. It is understood that at criticality the short-range local interactions between the two-state elements generate long-range global correlations, thereby producing effective long-range interactions. Consequently at criticality there is global cooperation among the network elements.

An individual disconnected from the network is assumed to choose randomly between two states with an exponential distribution of decision times and a given average decision time. When coupled to the other individuals of the network, the global distribution for the time intervals between decisions is determined to be inverse power-law \cite{west13}. In this latter case the power-law index of the survival probability is a measure of the complexity of the underlying dynamics determining whether that process is non-stationary and non-ergodic \cite{turalska09,turalska11}. In Section \ref{subordination} the DMM network dynamics is incorporated into that of an individual element through a process known as subordination. In order to formalize the subordination process we introduce the concept of subjective time to distinguish between clock time that determines the activities of the network and operational or subjective time that determines the activities of the individual.

The subordination process results in the two-state master equation of the DMM being replaced by a fractional master equation for the individual whose solution is shown to be a Mittag-Leffler function in Section \ref{subordination}. This predicted behavior of the single element dynamics is compared with the numerical results from the DMM implemented on a two-dimensional lattice and found to be in excellent agreement. In Section \ref{conclusion} we draw some conclusions.

\section{Decision Making Model (DMM)\label{DMM}}

The DMM realized on a complex networks represents the dynamics of the probability for an individual to be in either of the two states: yes or no, up or down, on or off. The model is based on the cooperative interaction of $N$ elements, each of which is described by the two-state master equation \cite{turalska09,turalska11}

\begin{equation}
\frac{d}{dt}p_{1}^{(i)}=-g_{12}^{(i)}p_{1}^{(i)}+g_{21}^{(i)}p_{2}^{(i)},
\label{firstconsensus}
\end{equation}

\begin{equation}
\frac{d}{dt}p_{2}^{(i)}=-g_{21}^{(i)}p_{2}^{(i)}+g_{12}^{(i)}p_{1}^{(i)}.
\label{secondconsensus}
\end{equation}

The quantity $p_{j}^{(i)}\left( t\right) $ is the probability of the element \emph{i=1,...,N} in the network being in the state $j=1,2$ at time $t$ and the probability is normalized at all times such that

\begin{equation}
p^{(i)}_{1}(t)+p^{(i)}_{2}(t)=1.  \label{probability}
\end{equation}

The network dynamics are determined by the choice of the functional form of the transition rates in the two-state master equation (Eqs. (\ref{firstconsensus}) and (\ref{secondconsensus})). Each probability $p_{j}^{(i)}\left( t\right) $ is influenced by the states occupied by all the elements of the network linked to element \emph{i} as determined by the transition rates

\begin{equation}
g_{12}^{(i)}(t)=g_{0}\exp \left[ K\left\{ \frac{N_{2}^{(i)}(t)-N_{1}^{(i)}(t)%
}{N^{(i)}}\right\} \right]   \label{transition12}
\end{equation}

\begin{equation}
g_{21}^{(i)}(t)=g_{0}\exp \left[ K\left\{ \frac{N_{1}^{(i)}(t)-N_{2}^{(i)}(t)%
}{N^{(i)}}\right\} \right]   \label{transition21}
\end{equation}

The symbol $N^{(i)}$ denotes the total number of elements linked to the \emph{i}-th element and $N_{s}^{(i)}(t)$ is the number of those elements in the state $s=1,2$ at time $t$. Of course $N^{(i)}=N_{1}^{(i)}+N_{2}^{(i)}$ at all times. The parameter $K$ is the control parameter that determines the strength of the interaction between elements of the network. In the case where each element in the network is coupled to all the other elements we have all-to-all (ATA) coupling, such that $N^{(i)}=N$ and the time dependence of the total number of elements in states $s=1,2$ $N_{s}^{(i)}(t)=N_{s}(t)$ implies that the transition rates become erratic functions of time.

\subsection{When every element is interconnected}

In the ATA coupling case when the total number of elements within the network becomes infinite ($N\longrightarrow \infty $) the fluctuation frequencies collapse into probabilities according to the law of large numbers. In physics this replacement goes by the name of \emph{the mean field approximation}, in which case the transition rates in the master equation (\ref{firstconsensus}) and (\ref{secondconsensus}) are written as

\begin{equation}
g_{12}(t)=g_{0}\exp\left[ -K \left\{ p_{1}(t)-p_{2}(t) \right\} \right]
\end{equation}

\begin{equation}
g_{21}(t)=g_{0}\exp\left[ -K \left\{ p_{2}(t)-p_{1}(t) \right\} \right]
\end{equation}

The formal manipulation of the master equation even in this simplified case in made a little easier if we introduce a new variable defined as the difference in the probabilities

\begin{equation}
\Pi(t) \equiv p_{1}(t)-p_{2}(t).  \label{difference}
\end{equation}

Subtracting Eq.(\ref{secondconsensus}) from Eq.(\ref{firstconsensus}) after some algebra yields the highly nonlinear rate equation for the difference variable

\begin{equation}
\frac{d}{dt}\Pi =-(g_{12}+g_{21})\Pi +(g_{21}-g_{12})  \label{notdefined}
\end{equation}

where the nonlinearity enters through the transition rate dependence on the difference variable

\begin{equation}
g_{12}=g_{0}\exp \left[ -K\Pi \right]   \label{model}
\end{equation}

\begin{equation}
g_{21}=g_{0}\exp \left[ K\Pi \right]   \label{model2}
\end{equation}

in the mean field approximation. By inserting Eqs.(\ref{model}) and (\ref{model2}) into Eq.(\ref{notdefined}) we obtain

\begin{equation}
\frac{d}{dt}\Pi =-\frac{\partial V}{\partial \Pi }  \label{ideal}
\end{equation}

and the network dynamics are determined by the potential function $V$($\Pi )$, which is a symmetric double well potential with the explicit form

\begin{equation}
V(\Pi )=\frac{2g_{0}}{K}\left[ \Pi \sinh K\Pi -\frac{K+1}{K}\cosh K\Pi \right] .  \label{potential}
\end{equation}

\begin{figure}[t]
\centering
\includegraphics[scale=1.50]{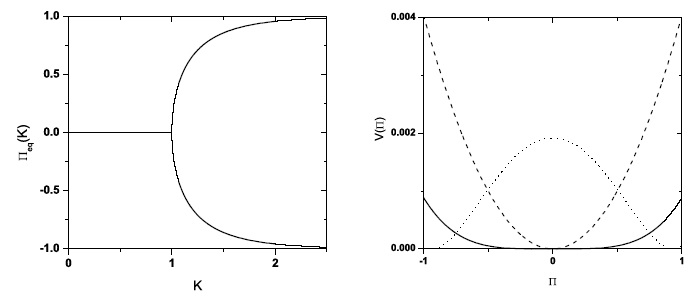}
\caption{ Left panel: The equilibrium mean field for different values of the control parameter $K$. A bifurcation occurs at the critical point $K=K_{c}=1.00.$ Right panel: Potential function determined by Eq.(\protect\ref{potential}) with barriers for K subcritical (dashed line, $K=0.20$), critical (solid line, $K=1.00$) and supercritical (dotted line, $K=1.80$).}
\label{fig_well}
\end{figure}

Note that the network is not described by a Hamiltonian and yet the global dynamics in indeed described by an effective Hamiltonian, that being the double well potential given by Eq.(\ref{potential}) and depicted in Figure \ref{fig_well}.

The cooperative behavior of the infinitely large ATA coupled network described by Eq.(\ref{ideal}) is that of an overdamped particle hopping from one potential minimum to the other, whose position is $\Pi $ within the potential Eq.(\ref{potential}). For $K<1$, half of the nodes are in one state and half are in the other because there is only a single broad minimum in the potential. At the critical value of the control parameter $K=K_{C}=1.0$ a bifurcation occurs and the potential develops two wells separated by a barrier as discussed by Turalska \textit{et al}. \cite{turalska09}. The height of the barrier increases with the value of the control parameter.

It is now convenient to define the stochastic global variable

\begin{equation}
\xi (t)=\frac{N_{1}(t)-N_{2}(t)}{N}=\frac{1}{N}\sum_{i=1}^{N}s_{i}(t),
\label{fluctuationxi}
\end{equation}

where $s_{i}(t)$ is the state of element $i$ at time $t$. The variability of the global variable characteristic of the entire network, capturing the cooperation between units at any moment of time. It is interesting that at the critical value of the control parameter the ATA version of the DMM undergoes a phase transition. Note that the amplitude of $\xi (t)$ depends on the value of the control parameter $K$. When $K=0$, all elements in the network are independent Poisson processes; thereby an average taken at any moment of time over all of them yields zero. Once the value of the coupling becomes nonzero, $K>0$, single elements are less and less independent, resulting in nonzero averages. The quantity $K_{c}$ is the critical value of the control parameter $K$, at which point a phase transition to a global majority state occurs.

\begin{figure}[t]
\centering
\includegraphics{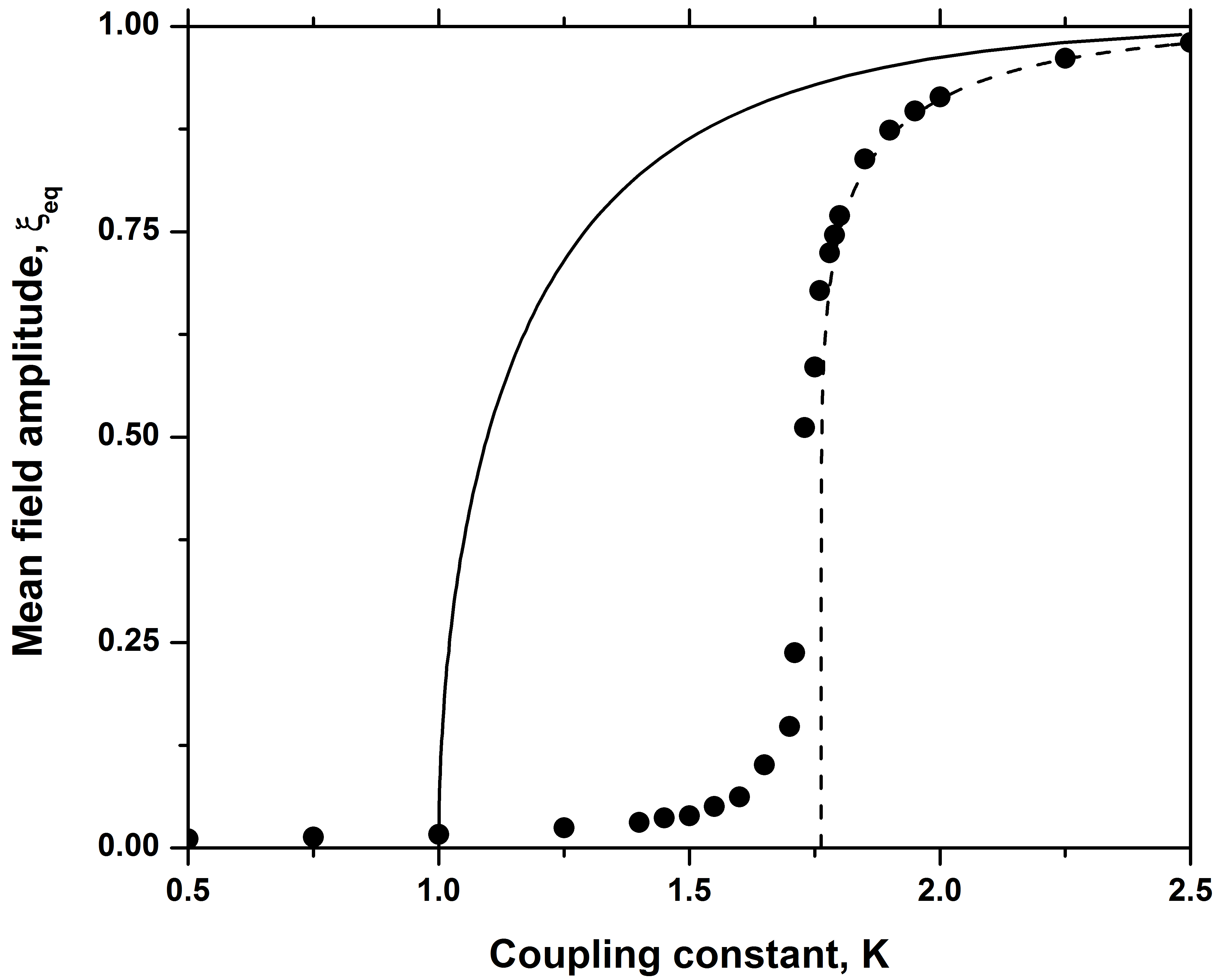}
\caption{ The phase diagram for the global variable $\protect\xi \left( t\right).$ The thin solid line and the dashed line are the theoretical predictions for the fully connected and the two-dimensional regular network, respectively. In both cases $N=\infty $ and the latter case is the Onsager theoretical prediction \protect\cite{onsager44} for a two-dimensional regular lattice. The dots corresponds to the global states observed for the DMM on a two-dimensional regular lattice $N=100\times 100$ nodes and $g_{0}=0.01$. Periodic boundary conditions were applied in the DMM calculations.}
\label{fig_phase}
\end{figure}

In numerical calculations we use the time average $\xi _{eq}$ $=$ $\left\langle \left\vert \xi \left( t\right) \right\vert \right\rangle $ as a measure of this global majority. More precisely, after an initial $10^{6}$ time steps, the average is taken over the same number of the consecutive time steps of the model. In Figure \ref{fig_phase} the thin line indicates the ATA phase transition as measured by $\xi _{eq}.$ The other phase transitions indicated are for the Ising model (dashed line) and the DMM model on a two-dimensional lattice as discussed in Section \ref{two_dim} and elsewhere \cite{turalska11}.

Real network are not ATA coupled since interactions typically have finite range and elements are spatially separated. Thus, the ATA approximation may be valid for small networks but certainly breaks down for large systems. Moreover, real-world networks have finite numbers of elements. It is therefore useful to examine how strongly the mean field solutions are violated when we relax these constraints. The stability condition can be violated in at least two different ways. The first way is by reducing the number of elements $N$ to a finite value. The second way is by restricting the number of links so the network no longer has ATA coupling. In real-world networks both sources of equilibrium disruption are expected to occur. For the time being we retain the ATA coupling within the networks and consider the number of elements $N$ to be finite. In this latter case we can no longer make the mean field approximation and the dynamic picture stemming from the above master equation is radically changed.

If the number of elements is still very large, but finite, we consider the mean-field approximation to be nearly valid and replace the average Eq.(\ref{fluctuationxi}) with the stochastic quantity

\begin{equation}
\xi (t)=\Pi (t)+f(t)  \label{generalization}
\end{equation}

where $f(t)$ is a small amplitude random fluctuation. After inserting Eq.(\ref{generalization}) into (\ref{ideal}) and some straight forward algebra we obtain the stochastic differential equation \cite{bianco08,grigolini13} to lowest-order in the strength of the fluctuations:

\begin{equation}
\frac{d\xi \left( t\right) }{dt}=-\frac{\partial V(\xi )}{\partial \xi }+\varepsilon \left( t\right) .  \label{stochastic ideal}
\end{equation}

The additive fluctuations $\varepsilon \left( t\right) $ have amplitudes that are computationally determined to be on the order of $1/\sqrt{N}.$

Note that the double-well potential in the mean field approximation persists in the present description even though we have relaxed the mean field approximation to a finite number of network elements. The random fluctuations resulting from the finite size of the network induces transitions between the two states of the potential well. Consequently, for a network with a finite but large number of elements the phase synchronization of Eq.(\ref{ideal}) is not stable and the stochastic Langevin equation Eq.(\ref{stochastic ideal}) determines the dynamics of the network. Furthermore, the fluctuations can drive the particle from one well of the potential to the other when its amplitude is sufficient to traverse the barrier separating the wells. However, here the fluctuations arise from the finite number of elements in the network rather than from non-existent thermal excitations and are consequently suppressed as the network size increases.

\begin{figure}[t]
\centering
\includegraphics[scale=1.00]{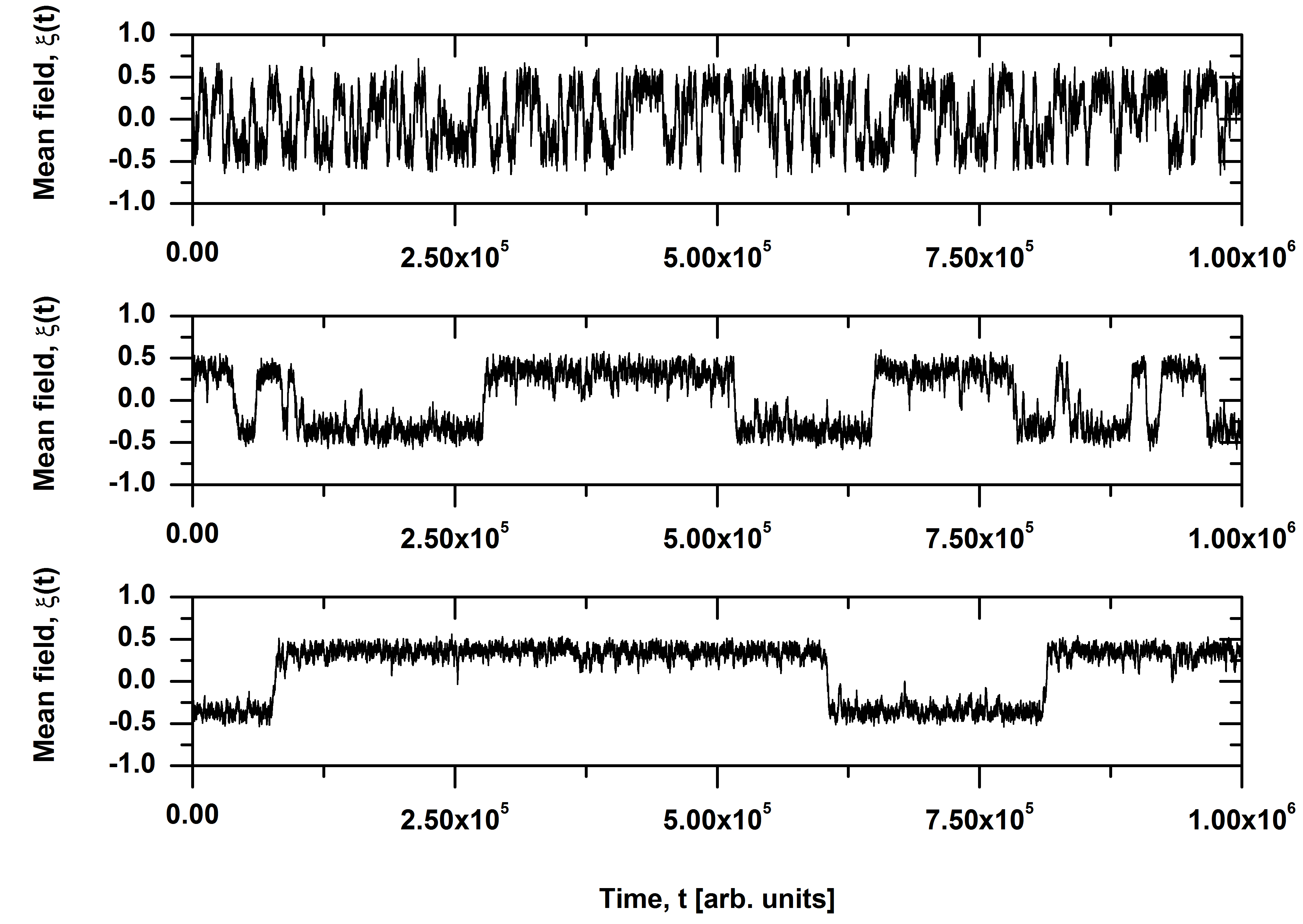}
\caption{ The fluctuation of the mean field-average phase as a function of time. For a system of $N=500$ elements (top), $N=1500$ elements (middle), and $N=2500$ elements (bottom). The coupling constant is $K=1.05$ and $g_{0}=0.01$ in all three cases.}
\label{fig_fluctuating}
\end{figure}

Although Eq. (\ref{stochastic ideal}) is written in the continuous time representation, in practice the numerical calculations of DMM correspond to the adoption of a finite integration time step $\Delta t=1$. Note that the stochastic rate equation Eq.(\ref{stochastic ideal}) replaces Eq. (\ref{ideal}) in the case of a finite $N$, and that Eq. (\ref{ideal}) is recovered in the ideal case $N=\infty $. We incorporate the ATA coupling condition with a finite number of elements by numerically integrating the the master equation for each element in the network and then calculating the number of elements in each of the two states. In Figure \ref{fig_fluctuating} the fluctuating global variable $\xi (t)$ is depicted as a function of time, under differing conditions. Notice that with increasing $N$ the fluctuation $\xi (t)$ become more distinctly dichotomous-like, with an increasingly sharp transition from the 'up' to the 'down' state. This pattern corresponds to the entire network keeping a decision for a longer and longer time as the
size of the network increases. The condition of a decision lasting forever is reached in the ideal case $N=\infty $. The global variable fluctuates between the two minima of the double-well potential as described by Eq.(\ref{stochastic ideal}) for $K=1.05>K_{C}$. The single element follows the fluctuations of the global variable, switching back and forth from the condition where the upper state is preferred statistically to that where the lower sate is preferred statistically. The complete properties of the DMM on an ATA network are explored by Turalska \emph{et al.} \cite{turalska09,turalska11}.

\subsection{Nearest neighbor coupling\label{two_dim}}

\begin{figure}[t]
\centering
\includegraphics[scale=1.00]{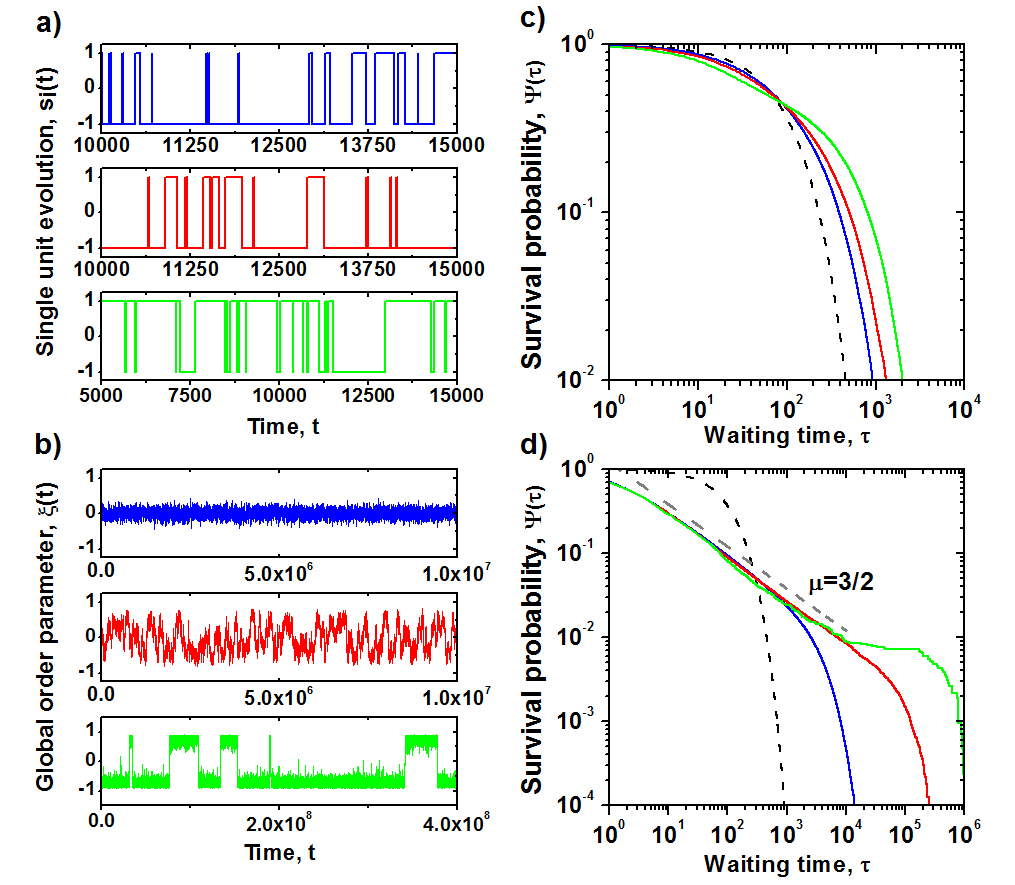}
\caption{ Behaviour of a discrete, two-state dynamic unit on a two-dimensional lattice. Temporal evolution and corresponding survival probability $\Psi \left( \protect\tau \right) $ for the transitions between two states for the single unit $(a,c)$ of the system is compared with the behavior and statistical properties of the global order parameter $(b,d)$. Simulations were performed on a lattice of size $N=50\times 50$ nodes, with periodic boundary conditions, for $g_{0}=0.01$ and increasing values of the control parameter $K$. Blue, red and green lines correspond to $K=1.50$, $1.70$ and $1.90$, respectively. The critical value of the control parameter is $K_{c}\approx 1.72.$ Black dashed line on the plots of $\Psi \left( \protect\tau \right) $ denotes an exponential distribution, with the decay rate $g_{0}$. The gray dashed line denotes an inverse power law function with exponent $\protect\mu -1.$ }
\label{fig_lattice}
\end{figure}

In this section we consider the topology of a simple two-dimensional lattice and confine the coupling between elements to its four nearest neighbors thereby setting $N=4$ in the transition rates of the two-state master equation. Similarly to the ATA case, the fluctuations of the global variable $\xi (t)$ show pronounced transition as a function of the coupling parameter $K$. As seen in Figure \ref{fig_lattice}b, the global variable shifts from a configuration dominated by randomness to an organized state once the control parameter is increased above the critical value $K_{C}$. For values of the control parameter $K$ corresponding to the disorganized phase $K<K_{C}$, single elements of the lattice are only weakly influenced by the decisions of the neighbours. Thus, the fluctuations of the global order parameter $\xi(t)$ are characterized by small amplitude and very fast oscillations about the zero-axis. For $K>K_{C}$, the interaction between individuals give rise to a majority or a consensus state, during which a significant number of agents adopts the same opinion at the same time.

At the same time the global behavior is undergoing a phase transition, the presence of the lattice apparently exerts only very subtle influence over the behavior of single individuals. The latter influence can be observed as a change in the interval timing for a single element as the control parameter is increased (Fig. \ref{fig_lattice}a). Note that if attention is concentrated on a single network element when the network is in a consensus state that individual would still appear to make transitions according to an exponential distribution as exhibited in Figure \ref{fig_lattice}c. The strict exponential is indicated by the black dotted curve. The single particle survival probabilities do not look too much different, the light gray dashed curve with the subcritical value $K=1.5<K_{c}$ is very close to the exponential. The remaining single particle curves, whether critical $K_{c}\approx 1.70$ or supercritical $K>K_{c}$ appear to be exponential on this graph.

To characterize the changes in temporal properties of the microscopic and macroscopic variables we evaluate the survival probability function $\Psi(\tau )$ of time intervals $\tau $ between consecutive events defined as changes of the state or crossing of the zero-axis, for the single element and the global variable, respectively. These calculations unveil modest deviation of the survival probability for a single individual from the exponential form $\Psi (\tau )=\exp [-g_{0}\tau ]$. The strict exponential is indicated in Figure \ref{fig_lattice}c by the black dashed curve. The single particle survival probabilities do not look too much different, the blue curve with the subcritical value $K=1.5<K_{C}$ is very close to the exponential. The remaining single particle curves, whether critical $K=K_{C}\approx 1.70$ or supercritical $K>K_{C}$ appear to be very nearly exponential on this graph. The difference in the behavior of the individual from that in the non-interacting state would be that she tends to be more reluctant to change her mind.

The deviation of the individual survival probability from the exponential form in Figure \ref{fig_lattice}c appears to be modest when compared with the dramatically greater deviation of the survival probability of the global variable from the exponential as depicted in Figure \ref{fig_lattice}d. The average network behavior differs markedly as the control parameter is increased from the subcritical through the supercritical regions. However the influence of the global variable on the behavior of the individual does not appear to induce a significant change. For the individual the change is however a subtle yet profound difference and is a direct result of the imitation mechanism, that is the ERH. So if the individual survival probability is not exponential, what is it? To answer this question we turn our attention to describing an alternate construction of the dynamics of the
individual elements.

\section{Subordination and Fractional Dynamics - two views of time\label{subordination}}

In this section we demonstrate the equivalence between a fractional trajectory that is the solution of a Caputo fractional differential equation, and the ensemble average trajectory that results from a subordination process. We here consider only fractional derivatives of the Caputo type, in part because it requires the least amount of explanation. However for the aficionado we note that an approach using Riemann-Liouville fractional derivatives would be equivalent as long as the initial conditions are properly specified. We begin the discussion with a derivation of the fractional derivative from a subordination argument.

The master equation for a single isolated individual is, with the index suppressed, according to the numerical simulation given by

\begin{equation}
\phi \left( n\Delta \tau \right) -\phi \left( \left[ n-1\right] \Delta \tau \right) =-g_{0}\phi \left( \left[ n-1\right] \Delta \tau \right) \Delta \tau ,  \label{singleME}
\end{equation}

whose discrete solution is

\begin{equation}
\phi \left( n\right) =\left( 1-g_{0}\Delta \tau \right) ^{n}\phi \left(0\right) .  \label{discrete}
\end{equation}

Here $\phi (n)$ is the difference variable for a single individual chosen from the network at random and as $n\rightarrow \infty $ and $\Delta \tau \rightarrow 0$ such that clock time is $t=n\Delta \tau $ we have the apparently trivial result

\begin{equation}
\phi \left( t\right) =e^{-g_{0}t}\phi \left( 0\right).  \label{phi_solution}
\end{equation}

Subordination implies the existence of two different notions of time \cite{sokolov06,svenkenson13}. One is the operational time $\tau $ , which is the internal time of a single individual, with an element generating the ordinary dynamics of a non-fractional system. The other notion is experimental time $t$; the time as measured by the clock of an external observer. Typically in the operational time frame the temporal behavior of an element is regular and evolves exactly according to the ticks of a clock leading to the exponential. Therefore it is assumed that the trajectory of a network's  element in operational time is well defined and given by $\phi(\tau )$, which is the solution given by Eq.(\ref{phi_solution}).

It is perhaps worthwhile to point out that this notion of two different times was introduced into psychology in the middle nineteenth century and lead to the general Weber-Fechner law. It has been further developed in a contemporary setting to explain the observation of $1/f$ noise in cognition by discriminating between subjective and objective times, that being operational and chronological time, respectively.

In operational time an element's behavior appears ordinary, but to an experimenter observing the elements from outside the network their temporal behavior appears erratic, evolving in time then abruptly freezing in different states for extended time periods. Because of the random nature of the experimental or chronological time evolution of the elements the subordination process involves an ensemble average over many realizations of the element's dynamics each evolving according to its own internal clock, independent of one another. Making an ensemble average over a large number of realizations results in a smooth average trajectory, which is equivalent to the fractional trajectory.

To find the average behavior we move from the operational time solution to the experimental time solution adopting the subordination interpretation. We assume that the chronological time lies in the interval $(n-1)\Delta \tau \leq t\leq n\Delta \tau$ and obtain

\begin{equation}
\langle \phi \left( t\right) \rangle = \sum^{\infty}_{n=1} \int^{t}_{0} \Psi(t-t') \psi_{n}(t') \phi(n)dt'.
\label{GME}
\end{equation}

It is evident that the trajectory resulting from the subordination process inherently involves an ensemble average. Here we see that Eq. \ref{GME} replaces the solution to the single element two-state master equation of the DMM.

Note that $\psi_{n}\left( t \right) dt$ is the probability that $n$ events have occurred, the last one in the time interval $(t,t+dt)$. The function $\Psi(t)$ denotes the probability that no event occurs up to time $t$ and is given empirically by numerical calculation in Figure \ref{fig_lattice}d and mathematically by Eq.(\ref{survival2}). The occurrence of an event corresponds to activating a decision with $(1-g_{0}\Delta \tau )$, so that activating $n$ such events transforms the initial condition $\phi(0)$ into the product $(1-g_{0}\Delta \tau )^{n}\phi \left( 0\right) $. This form of the equation is kept from time $t^{\prime}$, at which time the last event occurs, up to time $t$, the time interval $t-t^{\prime}$ being characterized by no event occurring. Of course, the expression Eq.(\ref{GME}) takes into account that the number of possible events may range from the
no-event case to that of infinitely many events. The conditions necessary for this result to occur are discussed by Svenkenson \emph{et al}\cite{svenkenson13}. To interpret the physical meaning of Eq. (\ref{GME}), consider each tick of the internal clock $n$ of an element measured in experimental time as an event. Since the observation is made in experimental time, the time intervals between events are random. We assume that the waiting times between consecutive events are identically distributed independent random variables. The integral in Eq. (\ref{GME}) is then built up according to renewal theory. After the $n$-th event the individual changes from state $\phi(n)$ to $\phi(n+1)$, where it remains until the action of the next event. The sum over $n$ takes into account the possibility that any number of events could have occurred prior to an observation at experimental time $t$. The events occur randomly with a waiting-time probability density function (\emph{pdf}) $\psi(t)$ and survival probability $\Psi(t)$. The waiting-time \emph{pdf} is related to the survival probability through

\begin{equation}
\psi (t)=-\frac{d\Psi(t)}{dt}
\end{equation}

Taking advantage of the renewal nature of the events, the waiting-time \emph{pdf} for the $n$-th event in a sequence is connected to the previous event by

\begin{equation}
\psi_{n}(t)=\int_{0}^{t}\psi_{n-1}(t^{\prime})\psi(t-t^{\prime})dt^{\prime}  \label{psiN}
\end{equation}

At this point it is useful to introduce Laplace variables in our discussion. The Laplace transform of a function $f(t)$ is denoted

\begin{equation}
\widehat{f}(s)\equiv \int^{\infty}_{0} \exp^{-st}f(t)dt.
\label{transform}
\end{equation}

To find an analytical expression for the behavior in experimental time it is convenient to study the Laplace transform of Eq. (\ref{GME})

\begin{equation}
\langle \widehat{\phi }(s)\rangle =\widehat{\Psi }(s)\sum_{n=0}^{\infty}(1-g_{0}\Delta \tau )^{n}\left[ \widehat{\psi }(s)\right] ^{n}\phi (0).
\label{psi_hat}
\end{equation}

where we assume the intervals between successive transitions are independent of one another, it is a renewal process. Consequently, the waiting time \emph{pdf} for $n$ transitions is the product of $n$ single transition \emph{pdf}'s:

\begin{equation}
\widehat{\psi}_{n}\left( s\right) =\left[ \widehat{\psi }\left( s\right) \right] ^{n}
\end{equation}

which was used to collapse the convolution of Eq.(\ref{psiN}) to the power-law form in Eq.(\ref{psi_hat}).

Consequently the time $t$ is derived from a waiting-time $pdf$ given by that of the network as a whole:

\begin{equation}
\psi \left( t\right) =\frac{\left( \mu-1\right) T^{\mu-1}}{\left(T+t\right) ^{\mu }}  \label{survival}
\end{equation}

and the survival probability is empirically determined from Figure \ref{fig_lattice}d to be

\begin{equation}
\Psi \left( t\right) = \int^{\infty}_{t}\psi \left(t^{\prime }\right) dt^{\prime }=\left( \frac{T}{T+t}\right)^{\mu -1}.
\label{survival2}
\end{equation}

The Laplace transform of the survival probability in terms of that for the waiting-time \emph{pdf} is

\begin{equation}
\widehat{\Psi }\left( s\right) =\frac{1}{s}\left[ 1-\widehat{\psi }\left(s\right) \right] .  \label{phi_last}
\end{equation}

Inserting these last two expressions into Eq.(\ref{psi_hat}) and evaluating the sum yields

\begin{equation}
\left\langle \widehat{\phi }(s)\right\rangle =\frac{1}{s+g_{0}\Delta \tau\widehat{\Phi }\left( s\right) }\phi \left( 0\right)   \label{last}
\end{equation}

whose inverse Laplace transform yields:

\begin{equation}
\frac{d\langle \phi (t)\rangle }{dt}=-g_{0}\Delta \tau \int \Phi (t-t^{\prime })\langle \phi (t^{\prime })\rangle dt^{\prime }  \label{GME2}
\end{equation}

a generalized master equation.

\subsection{Fractional Langevin Equation \label{fractional_dynamics}}

The function $\Phi \left( t\right) $ in the Eq.(\ref{GME2}) is a memory kernel containing the information on how the other elements in the network influence the dynamics of the individual element under study. The Laplace transform of the memory kernel is

\begin{equation}
\widehat{\Phi }\left( s\right) =\frac{s\widehat{\psi }\left( s\right) }{1-\widehat{\psi }\left( s\right) }  \label{memory}
\end{equation}

and a complete discussion of its properties is now given in textbooks \cite{west11}. Equation (\ref{memory}) is the Laplace transform of the Montroll-Weiss memory kernel obtained using their continuous time random walk theory .

Previous analysis, including the DMM calculations, have shown that the global waiting-time $pdf$ is an inverse power-law distribution, see for example Figure \ref{fig_lattice}d. The asymptotic behavior of an individual in time is determined by considering the waiting-time $pdf$ given by Eq.(\ref{survival}) as $s\longrightarrow 0:$

\begin{equation}
\widehat{\psi }\left( s \right) \approx 1-\Gamma(2-\mu )\left( sT\right)^{\mu -1}; 1<\lambda <2
\label{expansion}
\end{equation}

so that Eq.(\ref{last}) reduces to

\begin{equation}
\widehat{\phi }\left( s\right) =\frac{1}{s+\lambda ^{\mu -1}s^{2-\mu }}\phi\left( 0\right)   \label{last2}
\end{equation}

and the rate parameter has the value

\begin{equation}
\lambda ^{\mu -1}=\frac{g_{0}\Delta \tau }{\Gamma \left( 2-\mu \right)T^{\mu-1}}.  \label{parameter}
\end{equation}

We now assume that the exact equation for the individual dynamics has both an average and a fluctuating part just as in the mean field treatment of the double well potential. Consequently in terms of the Laplace variables we have the stochastic equation

\begin{equation}
\widehat{\phi }\left( s\right) =\frac{1}{s+\lambda ^{\mu -1}s^{2-\mu }}\phi\left( 0\right) +\frac{1}{s+\lambda ^{\mu -1}s^{2-\mu }}\widehat{\varepsilon}\left( s\right)
\label{noise}
\end{equation}

which has the inverse Laplace transform \cite{west03}

\begin{equation}
\partial_{t}^{\mu -1}\left[ \phi (t)\right] =-\lambda ^{\mu -1}\phi(t)+\varepsilon (t).  \label{fractitonal LE}
\end{equation}

Equation (\ref{fractitonal LE}) is a stochastic fractional master equation or fractional Langevin equation in which the stochastic properties of $\varepsilon (t)$ are determined by the fluctuations resulting from the dynamics if the finite-size DMM network. The fractional derivative in this equation is of the Caputo form and has the Laplace transform

\begin{equation}
LT\left\{ \partial _{t}^{\alpha }[\phi (t)];s\right\} =s^{\alpha }\widehat{\phi }(s)-s^{\alpha -1}\phi (0)
\end{equation}

and is completely equivalent to that determined using the Riemann-Liouville form of the fractional derivative.

The solution to the fractional Langevin equation is given by

\begin{eqnarray}
\phi \left( t\right) &=&\phi \left(0\right) E_{\mu -1}\left( -\left( \lambda t\right) ^{\mu -1}\right)
\\
&&+\int^{t}_{0}\left( t-t^{\prime }\right) ^{\mu-2}E_{\mu -1,\mu -1}\left( -\left( \lambda \left[ t-t^{\prime }\right]\right) ^{\mu -1}\right) \varepsilon \left( t^{\prime}\right) dt^{\prime }.
\label{solution2}
\end{eqnarray}

where the homogeneous solution to the fractional equation is the Mittag-Leffler function

\begin{equation}
E_{\theta }\left( z\right) = \sum^{\infty}_{k=0}\frac{z^{k}}{\Gamma \left( 1+k\theta \right) }; \theta >0.
\label{MLF}
\end{equation}

and the kernel of the integral is in terms of the Mittag-Leffler function of the second kind

\begin{equation}
E_{\theta,\eta }\left( z\right) =\sum^{\infty}_{k=0}\frac{z^{k}}{\Gamma \left( \eta+k\theta \right) }.  \label{MLF2}
\end{equation}

The dynamics of the individual is determined by the exact Laplace transform equation Eq.(\ref{phi_last}). However it is notoriously difficult to obtain analytic expressions by direct inversion of the resulting equations due to the complexity of the exact form of the Laplace transform memory kernel. Consequently, the strategy is to consider the asymptotic forms of the solution, which was done by examining the solutions to the fractional Langevin equation given by Eq.(\ref{solution2}). We find that the properties of the fluctuations change as the control parameter is varied from the subcritical, critical and supercritical regions.

\subsection{Solution domains}

\begin{figure*}[t]
\centering
\includegraphics[scale=1.00]{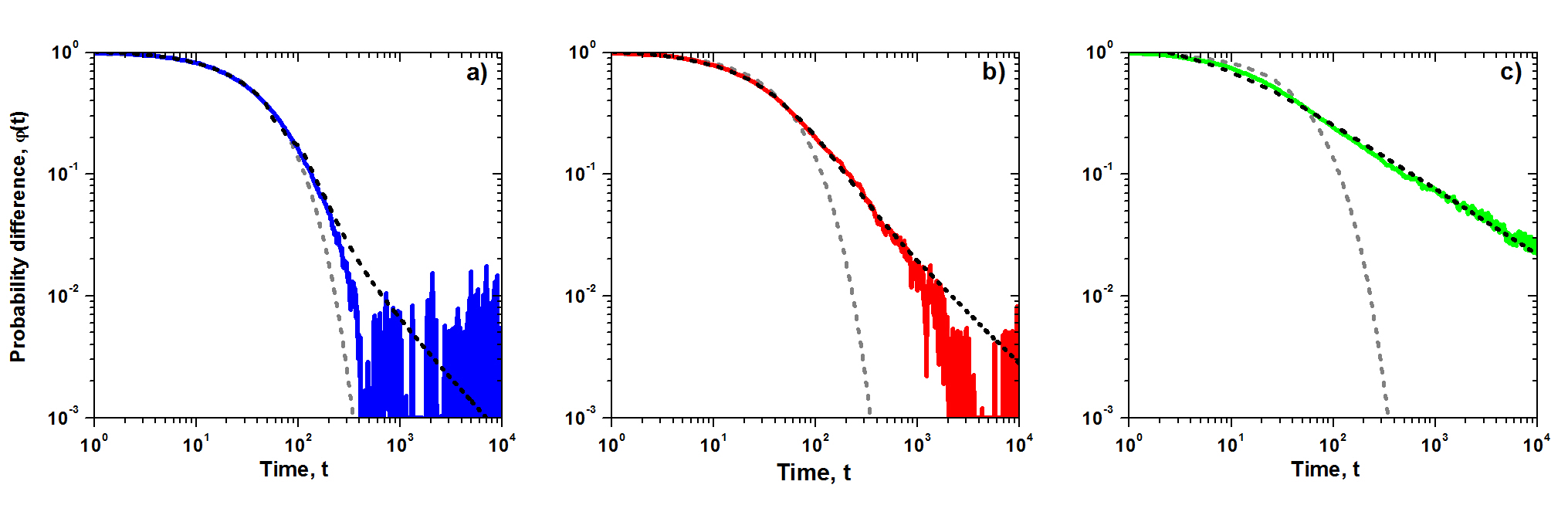}
\caption{ The probability difference $\left\langle \protect\varphi (t)\right\rangle $ estimated as an average over an ensemble of $10^4$ independent realizations of single element trajectories. Each trajectory corresponds to evolution of a randomly selected node on $N=100\times 100$ lattice with $g_{0}=0.01$ and the same initial condition $s_{i}(0)=1$: (a) subcritical domain, $K=1.00$; (b) critical domain, $K=1.70$; (c) supercritical domain, $K=2.50$. Grey dashed line denotes the exponential form of probability difference, $\protect\varphi (t)=exp(-2g_{0}t)$ that is obtained for a single isolated individual. Black dashed line denotes a fit with the Mittag-Leffler function (Eq.(\protect\ref{MLF})). }
\label{fig_difference}
\end{figure*}

In all three regions of DMM dynamics, subcritical, critical and supercritical, the single elements used in the evaluation of the probability difference $\left\langle \varphi (t)\right\rangle $ were selected at random among all nodes of the lattice. The calculations were done on a $100\times 100$ node two-dimensional lattice, with nearest neighbor interactions. The time-dependent average solution calculated over an ensemble of randomly chosen individuals is depicted in Figure \ref{fig_difference}, where the average is taken over $10^{4}$ independent realizations of the dynamics. The analytic solution is obtained by averaging Eq.(\ref{solution2}) over an ensemble of realizations of the single particle trajectory to obtain the Mittag-Leffler function:

\begin{equation}
\langle\phi \left( t\right)\rangle =\phi \left( 0\right) E_{\mu -1}\left( -\left[ \lambda t\right] ^{\mu -1}\right)  \label{MLFsolution}
\end{equation}

From the series form of the Mittag-Leffler function it is evident that for $\mu =2$ the average probability difference would be an exponential. Consequently the influence of the network on the behavior of the individual in this case would be essentially that of uncorrelated random noise and therefore would not qualitatively change from the Poisson nature of an isolated individual. However this is not the case for other values of the inverse power-law index; in the subcritical region the fitted value of the scaling index is $\mu =1.914$. Thus, the dynamic behavior of the network results in a stretched exponential autocorrelation for the dynamics of the individual

\begin{equation}
\lim_{t\longrightarrow 0} \langle \phi(t) \rangle =1-\frac{\left(\lambda t\right) ^{\mu -1}}{\Gamma \left( \mu \right) }\approx \exp \left( -\frac{\left[ \lambda t\right] ^{\mu -1}}{\Gamma \left( \mu \right) }\right) .
\label{stretched}
\end{equation}

Note that the early time behavior of the Mittag-Leffler function is the indicated stretched exponential. On the Figure \ref{fig_difference}a, the region where the black dashes of the Mittag-Leffler function fit diverge from the data is the onset of the inverse power-law tail of the Mittag-Leffler function. An exponential truncation of the Mittag-Leffler function would fit the data throughout its domain. The rational for a truncated Mittag-Leffler function will be taken up elsewhere. The fitting of the analytic solution at early times to the DMM numerically generated curves is certainly very good in the subcritical domain with $R^{2}=0.9968$.

As the critical point is approached from the subcritical region the random influence of fluctuations are diminished as would be expected due to the formation of clusters as the network undergoes a phase transition and encounters critical slowing down. The plunging stretched exponential that was observed in the subcritical region as seen in Figure \ref{difference}a is replaced with a more gently decreasing function. The time-dependent average solution of a randomly chosen individual in the critical regime is depicted in Figure \ref{difference}b. It is evident by comparing these data with the curve in Figure \ref{difference}a that the average solution does not decrease as quickly in time and there is less variability asymptotically in time. This behavior is reflected in the value of the power-law index which is determined to be $\mu =1.808$ with a quality of fit given by $R^{2}=0.9989$. Note how well separated the solution is from the exponential function given by the light grey dashed curve. But here again an exponential truncation of the Mittag-Leffler function might provide a better overall fit to the data.

In the supercritical region of the control parameter it is evident from the fit of the analytic solution to the data depicted in Figure \ref{difference}c that the solution extends far beyond that found in either the subcritical or critical domains with the Mittag-Leffler function solution extending far into the inverse power-law region. Here the power-law index is fitted with the value $\mu =1.534$ with $R^{2}=0.989$. The measured inverse power-law index is very close to that obtained for the global survival probability obtained from the numerical calculation of the DMM lattice network.

\section{Conclusion\label{conclusion}}

In summary the last few years have witnessed substantial attention focused on the role of criticality \cite{mora11} to explain the function of complex networks, from flocks of birds \cite{cavagna10}, to neural networks \cite{friedman12} to the brain \cite{chialvo10}. At criticality, the short-range links of Ising-like cooperative models are converted into long-range interactions turning a set of $N$ distinct elements into an organized network behaving as a single individual with extended cognition \cite{vanni11,bellomo12}. A complex network at criticality generates 1/f noise \cite{koverda11}, which is thought to be of relevance for cognition \cite{wagenmakers12}, with the interesting property of maximizing information transport \cite{arcangelis12,west11}. Moreover the network dynamics has a subtle but profound influence on the behavior of each individual within the network.

The numerical solution of the DMM on a $100\times 100$ lattice with elements at each of the nodes and with nearest neighbor interactions gives rise to an inverse power-law survival probability for the global variable introduced in Section \ref{DMM}. Using the theory of subordination, that being the time experienced by an individual, with the influence of the network entering into the individual's dynamics through the distribution of critical events, the dynamics of an individual is determined by a fractional Langevin equation.

The explicit form of the fractional Langevin equation depends on whether the network dynamics is in the subcritical, critical or supercritical domains. In all three domains the Mittag-Leffler function solution to the fractional Langevin equation is sufficient to describe the dynamic response of an individual to the other $9,999$ dynamic elements of the network. In the subcritical and critical domains the solutions could be modified to include truncations effects evident in the numerical data.

The lesson to be learned from the combination of computation and analysis presented herein is that complex networks of finite size whose dynamics are members of the Ising universality class, such as described by the DMM, have an analytic not just a numerical description. Instead of confining the dynamic description to that of the macroscopic variable, that being the global or average state of the network, one can also investigate how individual members of the network respond to the influence of the network as a whole. If we consider the fluctuations in the global dynamics to be microscopic, and the potential of the global variable to be macroscopic, then the real-time dynamic description of the individuals is mesoscopic. In general the mesoscopic dynamics can be described by a fractional stochastic differential equation.

Coupling two or more such fractional stochastic equations to model the across-scale coupling within the brain depicted in Figure \ref{fig_markram} suggests itself. This is presently an active area of investigation.

\section{Acknowledgement}

The authors would like to thank the U.S. Army Research Office for supporting
this research. P.G. warmly thanks the ARO and the Welch Foundation for their support through Grants No. W911NF-11-1-0478 and No.B-1577, respectively.

\end{document}